\documentclass[aps,pra,twocolumn,nofootinbib,floatfix,superscriptaddress]{revtex4-1}

\usepackage{amsmath,mleftright,mathtools}
\usepackage{bm}
\usepackage{stix,microtype}
\usepackage{graphicx}
\usepackage{xspace}
\usepackage{units}
\usepackage[colorlinks,linkcolor=blue,citecolor=blue,urlcolor=blue,breaklinks]{hyperref}
\usepackage[dvipsnames]{xcolor}
\usepackage{cleveref}
\usepackage{array}   
\newcolumntype{L}{>{$}c<{$}} 
\newcolumntype{C}{>{$}c<{$}} 
\usepackage{dcolumn}
\DeclareRobustCommand{\atez}{\text{\reflectbox{$z$}}}

\newcommand{\be}{\begin{equation}}
\newcommand{\ee}{\end{equation}}
\newcommand{\bea}{\begin{eqnarray}}
\newcommand{\eea}{\end{eqnarray}}

\def\b{\beta}
\def\g{\gamma}
\def\G{\Gamma}
\def\d{\delta}
\def\D{\Delta}

\def\th{\theta}
\def\Th{\Theta}

\def\m{\mu}

\def\r{\rho}

\def\S{\Sigma}
\def\t{\tau}

\def\Q{\Psi}

\def\callT{\mbox{$\mathcal{T}$}}

\def\iif{\infty}

\def\Tr{{\rm Tr}}

\def\1op{\hat{\mathbbm{1}}}
\def\nn{\nonumber}

\begin{document}
\title{Kadanoff-Baym equations for interacting systems with dissipative Lindbladian dynamics}
\author{Gianluca Stefanucci}
\affiliation{Dipartimento di Fisica, Universit{\`a} di Roma Tor Vergata, Via della Ricerca Scientifica 1,
00133 Rome, Italy}
\affiliation{INFN, Sezione di Roma Tor Vergata, Via della Ricerca Scientifica 1, 00133 Rome, Italy}

\begin{abstract}  	
The extraordinary quantum properties of nonequilibrium systems 
governed by dissipative dynamics have become a focal point in 
contemporary scientific inquiry. 
The Nonequilibrium Green's Functions (NEGF) theory provides a 
versatile method for addressing driven {\em non-dissipative} systems, 
utilizing the powerful diagrammatic technique to incorporate 
correlation effects. We here present a second-quantization approach to the
{\em dissipative} NEGF theory, 
reformulating Keldysh ideas to accommodate Lindbladian 
dynamics and extending the Kadanoff-Baym equations accordingly. Generalizing 
diagrammatic perturbation theory for many-body Lindblad operators, 
the formalism enables correlated and 
dissipative real-time simulations for the exploration of
transient and steady-state 
changes in the electronic, transport, and 
optical properties of materials.

\end{abstract}
\maketitle

Nonequilibrium systems governed by dissipative dynamics  
have broken into modern science owing to their remarkable quantum properties. 
Optical cavities of atoms~\cite{ritsch_cold_2013,baumann_dicke_2010}, 
molecules~\cite{flick_atoms_2017,mandal_theoretical_2023} 
and solid state 
systems~\cite{huebener_engineering_2021,schlawin_cavity_2022} offer 
an ideal platform to exploit the interplay between coherence and 
dissipation~\cite{verstraete_quantum_2009}. 
Special attention has thus far been dedicated to 
stationary states -- in particular the study of non-equilibrium fixed 
points and critical exponents~\cite{sieberer_dynamical_2013,kulkarni_cavity-mediated_2013}, 
phase 
transitions~\cite{dallatorre_quantum_2010,raftery_observation_2014}, 
entanglement~\cite{krauter_entanglement_2011,kastoryano_dissipative_2011}, 
and topology~\cite{bergholtz_exceptional_2021} -- as well as Floquet 
states~\cite{sato_floquet_2020,takashi_floquet_2023}. 

The transient dynamics of dissipative systems 
subjected to ultrafast driving fields remains 
poorly explored. In fact, the concomitant action of external fields, correlation and 
dissipation calls for innovative many-body frameworks. 
The Lindblad equation~\cite{lindblad_on_the_generators_1976,Breuer_the_theory_2007} serves as a solid 
ground to incorporate the aforementioned physics, preserving the trace and 
positivity of the many-body density matrix $\hat{\rho}$. However, 
its {\em brute force} numerical solution scales exponentially with 
the system size. Nonequilibrium Green's functions (NEGF) 
theory~\cite{svl-book,kamenev_field_2011,jauho-book} 
has proven to be a versatile tool to deal with driven systems; it  
leverages the powerful diagrammatic technique to account for  
correlation effects, thus reducing from 
exponential to power-law the numerical scaling. 
The inclusion of Lindblad dissipation in NEGF 
has been accomplished by Sieberer et al. in the so called  
field theory 
approach~\cite{sieberer_keldysh_2016,fogedby_field-theoretical_2022,thomson_field_2023,sieberer_universality_2023},
which is based on the path integral technique and Schwinger-Keldysh 
action~\cite{Altland_Simons_2010}.
Alternatively,  NEGF can be formulated in the second-quantization 
approach~\cite{rammer_quantum_2007,svl-book,balzer_nonequilibrium_2013,stefanucci_in-and-out_2023},
where concepts like the Martin-Schwinger hierarchy~\cite{martin_theory_1959},
Kadanoff-Baym equations~\cite{kb-book,danielewicz_quantum_1984}, 
conserving approximations~\cite{baym_conservation_1961,baym_self-consistent_1962} 
and Bethe-Salpeter equation~\cite{s.1986,sander_beyond_2015}  
are employed to develop many-body schemes for the 
simulation of 
carrier~\cite{steinhoff_nonequilibrium_2016,MolinaSanchez2017,perfetto_real_2022} and 
phonon~\cite{tong_toward_2021,caruso_nonequilibrium_2021,caruso_ultrafast_2022}
dynamics, dephasing and 
thermalization~\cite{perfetto_real-time_2023}, transient 
photoabsorbtion~\cite{attaccalite_real-time_2011,jiang_real_2021,PSMS.2015,tuovinen_comparing_2020}, 
photoemission~\cite{Freericks_theoretical_2009,PSMS.2016} and 
Raman~\cite{galperin_raman_2009,werner_nonequilibrium_2023} 
spectroscopy, 
photoexcitations and quenches in Mott and excitonic
insulators~\cite{aoki_nonequilibrium_2014,eckstein_thermalization_2009,MurakamiPRL2017,schuler_nessi_2020}, or  
time-dependent quantum 
transport~\cite{meir_landauer_1992,jauho_time_dependent_1994,galperin_molecular_2007,myohanen_a_many_body_2008,myohanen_kadanoff_baym_2009,pva.2010,schwarz_lindblad-drivem_2016}. 
The question of how the second-quantization 
approach -- and related concepts -- should be extended to encompass  
dissipation remains to be elucidated. 

In this work we offer a second-quantization perspective on the 
dissipative NEGF theory. We reformulate the original Keldysh 
idea to accommodate Lindbladian time-evolutions, and extend the 
Kadanoff-Baym equations accordingly. We also show how to generalize 
the diagrammatic perturbation theory for many-body Lindblad 
operators. The resulting formalism paves the way for real-time
simulations of the correlated and dissipative dynamics of 
materials.

{\em Keldysh-Lindblad formalism.--}
For systems with dissipative Lindbladian dynamics the average value 
of any, generally time-dependent, operator $\hat{O}(t)$ at time $t$ is expressed as 
$O(t)=\Tr\big[\hat{\r}(t)\hat{O}(t)\big]$, where the many-body density matrix 
satisfies the Lindblad equation (henceforth sum over repeated indices is 
implicit)
\begin{align}
\frac{d\hat{\r}}{dt}&=-i[\hat{H},\hat{\r}]_{-}
+2\hat{L}_{\g}\hat{\r}\hat{L}^{\dagger}_{\g}-
\hat{L}^{\dagger}_{\g}\hat{L}_{\g}\hat{\r}-
\hat{\r}\hat{L}^{\dagger}_{\g}\hat{L}_{\g}.
\label{lindbladeqp}
\end{align}
Here, $\hat{H}$ is the self-adjoint Hamiltonian of the system 
and $\hat{L}_{\g}$ are the Lindblad 
operators. Equation~(\ref{lindbladeqp}) can equivalently be cast in 
the form of an integral equation. Let 
$\hat{H}_{o}(t)= \hat{H}(t)-i\hat{L}^{\dagger}_{\g}(t)\hat{L}_{\g}(t)$ 
be the ``open system'' Hamiltonian and define the nonunitary 
evolution operator $\hat{U}_{o}(t,t')=T 
e^{-i\int_{t'}^{t}d\bar{t}\hat{H}_{o}(\bar{t})}$ for $t>t'$ and 
$\hat{U}_{o}(t,t')=\overline{T} e^{-i\int_{t'}^{t}d\bar{t}\hat{H}^{\dagger}_{o}(\bar{t})}$ 
for $t<t'$, where $T$ and $\overline{T}$ are the time and anti-time ordering 
operators, respectively. Then 
$\hat{\r}(t)=\hat{U}_{o}(t,0)\hat{\r}(0)\hat{U}_{o}(0,t)+
2\int_{0}^{t}dt_{1}\hat{U}_{o}(t,t_{1})
\hat{L}_{\g_{1}}(t_{1})\hat{\r}(t_{1})\hat{L}^{\dagger}_{\g_{1}}(t_{1})
\hat{U}_{o}(t_{1},t)$ solves Eq.~(\ref{lindbladeqp}), 
as it can easily be verified by direct 
differentiation. Iterating the integral equation and using the cyclic 
property of the trace, the time-dependent average $O(t)$ reads
\begin{align}
O(t)&=\Tr\Big[\hat{\r}(0)
\sum_{k=0}^{\iif}2^{k}\int_{0}^{t}dt_{1}..\int_{0}^{t_{k-1}}dt_{k}
\hat{U}_{o}(0,t_{k})L^{\dagger}_{\g_{k}}(t_{k})\ldots
\nn\\
&\times
\hat{U}_{o}(t_{2},t_{1})\hat{L}^{\dagger}_{\g_{1}}(t_{1})
\hat{U}_{o}(t_{1},t)\hat{O}(t)\hat{U}_{o}(t,t_{1})\hat{L}_{\g_{1}}(t_{1})
\hat{U}_{o}(t_{1},t_{2})
\nn\\
&\times 
\ldots L_{\g_{k}}(t_{k})\hat{U}_{o}(t_{k},0)\Big].
\label{aveOp}
\end{align}
This result can be written in a more useful form if we introduce the oriented 
Keldysh contour $C=C_{-}\cup C_{+}=(0,t)\cup (t,0)$. Let 
$z'\in C$ denote a contour-time and write $z'=t'_{\pm}$ if $z'\in 
C_{\pm}$. We also define the functions $\th_{\pm}(z')=1$ if $z'\in 
C_{\pm}$ and zero otherwise, and $s(z')=\th_{-}(z')-\th_{+}(z')$. Then 
\begin{align}
\hat{U}_{o}(t_{i},t_{j})=
\callT \Big\{e^{-i\int_{t_{j\mp}}^{t_{i\mp}}d\bar{z}
\big[\hat{H}(\bar{z})-is(\bar{z})
\hat{L}^{\dagger}_{\g}(\bar{z})\hat{L}_{\g}(\bar{z})\big]}\Big\}, \quad 
t_{i}\gtrless t_{j},
\end{align}
where $\callT$ is the contour ordering operator. If we
set the times of all operators $\hat{L}^{\dagger}$ 
on the $C_{+}$ branch and the times of all operators $\hat{L}$ 
on the $C_{-}$ branch then the string of operators in Eq.~(\ref{aveOp}) 
is contour ordered. 
Therefore, taking into account 
that under the $\callT$-sign the bosonic (fermionic) 
operators (anti)commute, we can 
extend all integration limits to $t$ and divide by $k!$. The 
re-ordering does not generate minus signs in the case of fermionic 
Lindblad operators since the number of interchanges 
is always even.
In this way the series is transformed into the Taylor expansion of an exponential, and 
the time-dependent average simplifies to
\begin{align}
O(t)=\Tr\Big[\hat{\r}(0)\callT\Big\{e^{-i\int_{C}d\bar{z}
\hat{H}(\bar{z},\bar{\atez})
}\hat{O}(z)\Big\}\Big],
\label{aveO3p}
\end{align}	
where 
\begin{align}
\hat{H}(\bar{z},\bar{\atez})=\hat{H}(\bar{z})-is(\bar{z})
\hat{L}^{\dagger}_{\g}(\bar{z})\hat{L}_{\g}(\bar{z})+2i\th_{-}(\bar{z})
\hat{L}^{\dagger}_{\g}(\bar{\atez})\hat{L}_{\g}(\bar{z}),
\label{hzatez}
\end{align}
and $\bar{\atez}=\bar{t}_{\mp}$ if $\bar{z}=\bar{t}_{\pm}$. Notice 
that under integration the last term in 
Eq.~(\ref{hzatez}) can alternatively be written as $-2i\th_{+}(\bar{z})
\hat{L}^{\dagger}_{\g}(\bar{z})\hat{L}_{\g}(\bar{\atez})$ since 
$\int_{C}d\bar{z} F(\bar{z},\bar{\atez})=-\int_{C}d\bar{z} F(\bar{\atez},\bar{z})$ 
for any function $F$, and $\th_{\pm}(\bar{z})=\th_{\mp}(\bar{\atez})$.
We also observe that in 
Eq.~(\ref{aveO3p}) the contour $C$ can be extended to infinity, i.e., 
$C=(0,\iif)\cup(\iif,0)$, since 
$\hat{K}\equiv \callT\Big\{e^{-i\int_{t_{-}}^{t_{+}}d\bar{z}
\hat{H}(\bar{z},\bar{\atez})}\Big\}=1$ for all 
$t$~\cite{notelk}.  This also implies that the contour-time $z$ of 
$\hat{O}$ can be either $t_{+}$ or $t_{-}$. 

In analogy with the theory of unitary evolutions we 
define the one-particle Keldysh-Lindblad NEGF according to 
\begin{align}
G_{ij}(z,z')=\frac{1}{i}\Tr\Big[\hat{\r}(0)
\callT\Big\{e^{-i\int_{C}d\bar{z}\hat{H}(\bar{z},\bar{\atez})
}\hat{d}_{i}(z)\hat{d}^{\dagger}_{j}(z')\Big\}\Big],
\label{klnegfdefmbp}
\end{align}	
where the annihilation operators $\hat{d}_{i}$ are either bosonic or 
fermionic, 
in which case $[\hat{d}_{i},\hat{d}^{\dagger}_{j}]_{\mp}=\d_{ij}$ (upper 
sign for bosons and lower sign for fermions). The contour argument of 
$\hat{d}$ and $\hat{d}^{\dagger}$ in Eq.~(\ref{klnegfdefmbp}) fixes the 
position of these operators along the contour, thus rendering 
unambiguous the action of $\callT$. The identity 
$\hat{K}=1$ implies that the NEGF satisfies the Keldysh properties
$G(t_{+},t'_{\pm})=G(t_{-},t'_{\pm})$ for 
$t>t'$ and $G(t_{\pm},t'_{+})=G(t_{\pm},t'_{-})$ for 
$t<t'$. 

Using the rules for the derivative of a contour-ordered string of 
operators, see  
Ref.~\cite{svl-book}, and introducing the shorthand notation 
$\left\langle\ldots\right\rangle\equiv\Tr\Big[\hat{\r}(0)
\callT\Big\{e^{-i\int_{C}d\bar{z}\hat{H}(\bar{z},\bar{\atez})
}\ldots\Big\}\Big]$,  we find the important result
\begin{align}
i\frac{d}{dz}G_{ij}(z,z')&=\frac{1}{i}\left\langle 
[\hat{d}_{i}(z),\hat{H}(z)-is(z)\hat{L}^{\dagger}_{\g}(z)\hat{L}_{\g}(z)]_{-}\hat{d}^{\dagger}_{j}(z')\right\rangle
\nn\\
-&2\th_{+}(z)\left\langle[\hat{d}_{i}(z),\hat{L}^{\dagger}_{\g}(z)]_{\mp}\hat{L}_{\g}(\atez)\hat{d}^{\dagger}_{j}(z')\right\rangle
\nn\\
\pm &2\th_{-}(z)\left\langle\hat{L}^{\dagger}_{\g}(\atez)
[\hat{d}_{i}(z),\hat{L}_{\g}(z)]_{\mp}\hat{d}^{\dagger}_{j}(z')\right\rangle
+\d(z,z'),
\label{klnegfderprelp}
\end{align}	
where the lower sign applies when {\em both} 
$\hat{d}_{i}$ and $\hat{L}_{\g}$ are fermionic 
operators, and the last term is the Dirac delta on the contour, i.e., 
$\int dz'\d(z,z')f(z')=f(z)$ for any function $f$. 
A similar equation can be derived by differentiating with 
respect to $z'$, see below and Appendix~\ref{eomapp}.
The (anti)commutators in 
Eq.~(\ref{klnegfderprelp}) generally give rise to higher-order NEGFs, 
the contour-time derivatives of which generate NEGFs of progressively 
higher order. In this manner,
the Martin-Schwinger hierarchy~\cite{martin_theory_1959} (MSH) for Lindbladian 
dynamics is established. 

For quadratic 
self-adjoint Hamiltonians 
$\hat{H}(t)=\hat{H}_{0}(t)=h_{mn}(t)\hat{d}^{\dagger}_{m}\hat{d}_{n}$ and 
linear Lindblad operators 
$\hat{L}_{1\g}(t)=a_{n}^{\g}(t)\,\hat{d}_{n}$ (one-body loss) 
and $\hat{L}_{2\g}(t)=b^{\g \ast}_{n}(t)\,\hat{d}^{\dagger}_{n}$ 
(one-body gain) the MSH couples the $n$-particle NEGF exclusively to 
the $(n-1)$-particle NEGF, and the solution of the MSH is equivalent 
to Wick's theorem~\cite{vanleeuwen_wick_2012}, see also Appendix~\ref{wickapp}. 
Equation~(\ref{klnegfderprelp}) and its analogous with derivative 
with respect to $z'$ reduce to (in matrix form)
\begin{subequations}
\begin{align}
\Big[i\frac{d}{dz}-\tilde{h}(z)\Big]G(z,z')+2i\ell(\atez)G(\atez,z')&=\d(z,z'),
\label{noninteom}
\\
G(z,z')\Big[\frac{1}{i}\frac{\overleftarrow{d}}{dz'}-\tilde{h}(z')\Big]
-2iG(z,\atez')\ell(z')
&=\d(z,z')
\label{klnegfderadjmbmgintp}
\end{align}	
\label{eomGkc}
\end{subequations}
where $\tilde{h}(z=t_{\pm})=h(t)-is(z)[\ell^{>}(t)\pm \ell^{<}(t)]$ and 
$\ell(z=t_{\pm})=\th_{-}(z)\ell^{>}(t)\mp \th_{+}(z)\ell^{<}(t)$, with  
$\ell^{>}_{mn}(t)=a^{\g\ast}_{m}(t)a^{\g}_{n}(t)$ and  
$\ell^{<}_{mn}(t)=b^{\g\ast}_{n}(t)b^{\g}_{m}(t)$ positive 
semidefinite and self-adjoint matrices. The time-dependence of $h(t)$ 
is generally due to external driving fields.
We refer to 
Eqs.~(\ref{eomGkc}) as the noninteracting dissipative equations of 
motion (eom).

{\em Interacting systems with one-body loss and gain.--}
In interacting systems the Hamiltonian 
$\hat{H}(t)=\hat{H}_{0}(t)+\hat{H}_{\rm int}(t)$, where 
$\hat{H}_{\rm int}$ is self-adjoint and 
at least quartic in the field operators. 
Expanding Eq.~(\ref{klnegfdefmbp}) in powers of $\hat{H}_{\rm int}$
and using Wick's theorem we obtain the Dyson equation (on the 
Keldysh contour)
$G=G_{0}+G_{0}\S G$, where $G_{0}$ satisfies Eqs.~(\ref{eomGkc})
and the many-body self-energy $\S$ is given by the sum of all one-particle 
irreducible Feynman diagrams. The interacting version of the dissipative eom 
follows when acting with $G_{0}^{-1}$ on the Dyson 
equation; the outcome is Eqs.~(\ref{eomGkc}) with a 
r.h.s. modified by the addition of  
$\int_{C}d\bar{z}\,\S(z,\bar{z})G(\bar{z},z')$ for 
Eq.~(\ref{noninteom}) and 
$\int_{C}d\bar{z}\,G(z,\bar{z})\S(\bar{z},z')$ for 
Eq.~(\ref{klnegfderadjmbmgintp}).

To derive the Kadanoff-Baym equations (KBE) 
satisfied by the lesser/greater 
NEGF $G^{\lessgtr}(t,t')\equiv G(t_{\mp},t'_{\pm})$, a 
preliminary 
discuss on the Langreth rules~\cite{l.1976} for convolutions on the Keldysh 
contour is needed. 
The many-body self-energy  has the structure~\cite{danielewicz_quantum_1984,svl-book,stefanucci_diagrammatic_2014} 
$\S_{ij}(z,z')=\d(z,z')V_{ij}(t)
+\left\langle 
\hat{\Q}_{i}(z)\hat{\Q}_{j}^{\dagger}(z')\right\rangle_{\rm irr}$, where 
$\hat{\Q}_{i}\equiv[\hat{d}_{i},\hat{H}_{\rm int}]_{-}$ and the subscript 
``irr'' signifies the irreducible part of the average. 
The quantity 
$V_{ij}(t)\equiv\left\langle[\hat{d}_{i}(z),\hat{\Q}_{j}^{\dagger}(z)]_{\mp}\right\rangle$
is the mean-field potential, hence the reminder is the 
correlation self-energy $\S^{\rm corr}$. 
The identity $\hat{K}=1$ guarantees that also 
$\S^{\rm corr}$ satisfies the Keldysh properties, i.e., 
$\S^{\rm corr}(t_{-},t'_{\pm})=\S^{\rm corr}(t_{+},t'_{\pm})$ for 
$t>t'$ and $\S^{\rm corr}(t_{\pm},t'_{-})=\S^{\rm corr}(t_{\pm},t'_{+})$ for $t<t'$. 
Therefore the Langreth rules are not affected by the Lindblad 
dissipators. We emphasize here the crucial role played by 
$L^{\dagger}_{\g}(\bar{\atez})\hat{L}_{\g}(\bar{z})$ in 
Eq.~(\ref{hzatez}). Excluding this term would leave us with two
different non-hermitian Hamiltonians, one on the forward branch and 
another on the backward branch, and 
hence $\hat{K}\neq 1$. The 
time-ordered and anti-time-ordered NEGF would then be independent 
functions, leading to
significantly more intricate Langreth rules~\cite{kantorovich_nonadiabatic_2018}. 
The KBE for $G^{<}$ is obtained by setting $z=t_{-}$ and $z'=t'_{+}$ 
in the interacting version of Eq.~(\ref{noninteom}). 
The second term on the l.h.s. yields the anti-time-ordered 
$G^{\rm {\bar T}}(t,t')\equiv G(t_{+},t'_{+})=G^{<}(t,t')-G^{\rm A}(t,t')$, 
where $G^{\rm A}(t,t')=-\th(t'-t)[G^{>}(t,t')-G^{<}(t,t')]$ is the 
advanced NEGF. Taking into account the 
definition of $\tilde{h}(z)$ and $\ell(z)$ we find
\begin{align}
\Big[i\frac{d}{dt}-
h_{o}(t)\Big]G^{<}(t,t')&\pm 2i\ell^{<}(t)
G^{\rm A}(t,t')
\nn\\
&=\big[
\S^{<}\cdot G^{\rm A}+
\S^{\rm R}\cdot G^{>}\big](t,t'),
\label{kb<tp}
\end{align}
where $h_{o}(t)\equiv h(t)-i\big(\ell^{>}(t)\mp \ell^{<}(t)\big)$ is the 
one-particle open-system Hamiltonian and the symbol ``$\cdot$'' is 
used for real-time convolutions between 0 and infinity.
As discussed in 
Refs.~\cite{dahlen_solving_2007,stan_time_2009,svl-book,schuler_nessi_2020} 
the anti-hermicity of $G^{\lessgtr}$ allows us to close the system of 
equations by solving the interacting version of Eq.~(\ref{klnegfderadjmbmgintp}) for $G^{>}$. 
For $z=t_{+}$ and $z'=t'_{-}$ the second term on the l.h.s. yields 
again $G^{\rm {\bar T}}$ which can also be written as $G^{>}-G^{\rm 
R}$, where $G^{\rm R}(t,t')=\th(t-t')[G^{>}(t,t')-G^{<}(t,t')]$ is the 
retarded NEGF. We then find
\begin{align}
G^{>}(t,t')\Big[\frac{1}{i}\frac{\overleftarrow{d}}{dt'}-
h^{\dagger}_{o}(t')
\Big]&+2iG^{\rm R}(t,t')\ell^{>}(t')
\nn\\
&=\big[
G^{>}\cdot \S^{\rm A}+
G^{\rm R} \cdot\S^{>}
\big](t,t').
\label{kb>tp}
\end{align}
Equations~(\ref{kb<tp}) and (\ref{kb>tp}) are the KBE for interacting 
dissipative systems with one-body loss and gain.
From the KBE it is straightforward to derive 
the eom for the retarded NEGF, i.e., 
$[i\frac{d}{dt}-h_{o}(t)]G^{\rm R}(t,t')=\d(t,t')
+\big[
\S^{\rm R} \cdot G^{\rm R}\big](t,t')$, as well as to show that 
$G^{\rm A}(t,t')=[G^{\rm R}(t',t)]^{\dagger}$.

{\em Lyapunov equation.--}
The eom for the one-particle 
density matrix $\r^{<}(t)=\pm iG^{<}(t,t)$ follows by subtracting 
Eq.~(\ref{kb<tp}) from its adjoint and then setting $t=t'$:
\begin{align}
\frac{d}{dt}\r^{<}(t)=-ih_{o}(t)\r^{<}(t)+i\r^{<}(t) h_{o}^{\dagger}(t)
+2\ell^{<}(t)+I(t),
\label{eomnfromGmg}
\end{align}
where $I(t)=\pm[\S^{<}\cdot G^{\rm A}+\S^{\rm R}\cdot 
G^{\rm <}](t,t)+{\rm h.c.}$ is known as collision integral. 
Notice that for $\S=0$, and hence $I=0$,
the solution of the KBE can be written as
\begin{align}
G^{\lessgtr}(t,t')=\pm G^{\rm R}(t,t')\r^{\lessgtr}(t')
\mp\r^{\lessgtr}(t)G^{\rm A}(t,t'),
\label{hfsol}
\end{align}
with $\r^{>}=\pm iG^{>}(t,t)=\r^{<}\pm 1$,
which is identical to the non-dissipative solution.

In the stationary case, i.e., $\frac{d}{dt}\r^{<}=0$, 
and for $I=0$ (no interaction) Eq.~(\ref{eomnfromGmg}) 
reduces to a Lyapunov equation, whose properties, e.g.,
 topological 
phases~\cite{lieu_tenfold_2020,atland_symmetry_2021,he_topological_2022}, 
exceptional points~\cite{wojcik_eigenvalue_2022,thomson_field_2023}, 
and bulk-edge 
correspondence~\cite{shen_topological_2018,yao_edge_2018,yokomizo_non-bloch_2019}, 
are currently under intense investigation, see also 
Refs.~\cite{sieberer_keldysh_2016,thomson_field_2023,sieberer_universality_2023} 
for an overview. 
A straightforward way to include correlation effects consists in 
evaluating the collision integral in the Boltzmann approximation, i.e., 
$I=-[(\G^{>}+\G^{<}),\r^{<}]_{+}+2\G^{<}$, 
where the 
rates $\G^{\lessgtr}[\r^{<}]$  for  in/out scatterings  
are functionals of $\r^{<}$~\cite{stefanucci_semiconductor_2023}. In this way, we 
revert to the 
noninteracting eom
with $\ell^{\lessgtr}\to \ell^{\lessgtr}+\G^{\lessgtr}$. 
In particular the stationary solution becomes a non-linear 
Lyapunov equation to be solved self-consistently. 

{\em Initial correlations.--}
As with any set of differential equations the KBE must be solved with 
an initial condition. For a system in thermal equilibrium before any 
external driving we have
$\hat{\r}(0)=\exp[-\b(\hat{H}-\m\hat{N})]/Z$, with $\b$ the inverse 
temperature, $\m$ the chemical potential, 
$\hat{N}=\sum_{i}\hat{n}_{i}$ the total number of particle operator 
and $Z$ the partition function. 
Initial correlations can be included 
by extending the Keldysh contour along the imaginary time axis until 
the point $-i\b$~\cite{kp.1961,Wagner_PhysRevB.44.6104}. This leads 
to a generalization of the KBE as the self-energy $\S(z,z')$ is 
nonvanishing for $z,z'\in(0,-i\b)$. Following 
Refs.~\cite{sa-1.2004,svl-book} we introduce the {\em left} 
$X^{\lceil}(\t,t)\equiv X(-i\t,t_{\pm})$ and {\em right}
$X^{\rceil}(t,\t)\equiv X(t_{\pm},-i\t)$ correlators with one real and 
one imaginary time argument (in $X^{\lceil}$ and $X^{\rceil}$ we can use either 
$t_{+}$ or $t_{-}$ since $\hat{K}=1$).   
Then Eqs.~(\ref{kb<tp}) and 
(\ref{kb>tp}) undergo modification with the inclusion of terms 
$[\S^{\rceil}\star G^{\lceil}]$ and $[G^{\rceil}\star \S^{\lceil}]$ 
on the r.h.s., respectively, where the symbol “$\star$” 
is used for imaginary-time convolutions between $0$ and $\b$.
Taking into account that 
$\tilde{h}(t_{\pm})-2i\ell(t_{\mp})=h_{o}(t)$ and 
$\tilde{h}(t_{\pm})+2i\ell(t_{\pm})=h^{\dagger}_{o}(t)$,
the interacting version of Eqs.~(\ref{eomGkc}) yield the eom 
for the left and right NEGF
\begin{subequations}
\begin{align}
\Big[i\frac{d}{dt}-
h_{o}(t)\Big]G^{\rceil}(t,\t)=\big[
\S^{\rceil}\cdot G^{\rm A}+
\S^{\rm M}\star G^{\rceil}\big](t,t'),
\label{kbrceiltp}
\\
G^{\lceil}(\t,t')\Big[\frac{1}{i}\frac{\overleftarrow{d}}{dt'}-
h^{\dagger}_{o}(t')
\Big]=\big[
G^{\lceil}\cdot \S^{\rm A}+
G^{\rm M} \cdot\S^{\lceil}
\big](t,t').
\label{kblceiltp}
\end{align}
\end{subequations}
The Matsubara correlators $X^{\rm M}(\t,\t')\equiv X(-i\t,-i\t')$ satisfy 
the Dyson equation $G^{\rm M}=G_{0}^{\rm 
M}+G_{0}^{\rm M}\star \S^{\rm M}\star G^{\rm M}$. This equation is 
decoupled from the left/right and lesser/greater NEGF,
and its solution provides the initial values 
for the KBE~\cite{dahlen_solving_2007,svl-book}.

{\em Two-particle loss.--} We say that the 
system is subject to a $n$-body loss (gain) if the Lindblad operator 
$\hat{L}_{\g}$ is a polynomial of order $n$ in the annihilation 
(creation) field operators. 
For $n>1$ the analytic solution is, in general, not available. 
We here show how to tackle the problem perturbatively by extending 
the many-body diagrammatic method. 
For the sake of definiteness we consider the set of Lindblad operators
$\hat{L}_{\g}=a^{\g}_{mn}\hat{d}_{m}\hat{d}_{n}$ 
(two-particle 
loss), with $a^{\g}_{mn}=\pm a^{\g}_{nm}$ for bosons/fermions. These dissipators are relevant in 
the context of exciton-polariton 
systems~\cite{carusotto_quantum_2013,wouters_excitations_2007}. 
Higher order loss and gain dissipators can be treated similarly. After some 
straightforward algebra the operator in Eq.~(\ref{hzatez}) is 
written as 
\begin{align}
\hat{H}(\bar{z},\bar{\atez})&=
\hat{H}(\bar{z})
\nn\\&-\frac{1}{2}
\int_{C}d\bar{z}'v^{\rm 2p}_{ijmn}(\bar{z},\bar{z}')
\hat{d}^{\dagger}_{i}(\bar{z}^{+})\hat{d}^{\dagger}_{j}(\bar{z}^{+})
\hat{d}_{m}(\bar{z}')\hat{d}_{n}(\bar{z}')
\label{Hzatez2bd}
\end{align}
where $v^{\rm 2p}_{ijmn}(z,z')=v^{\rm 2p}_{ijmn}(t)
[is(z)\d(z',z)+2i\th_{+}(z)\d(z',\atez)]$ and 
$v^{\rm 2p}_{ijmn}(t)
=2a^{\g\ast}_{ji}(t)a^{\g}_{mn}(t)=\pm v^{\rm 2p}_{jimn}(t)=
\pm v^{\rm 2p}_{ijnm}(t)$. 

\begin{figure}[tbp]
    \centering
\includegraphics[width=0.48\textwidth]{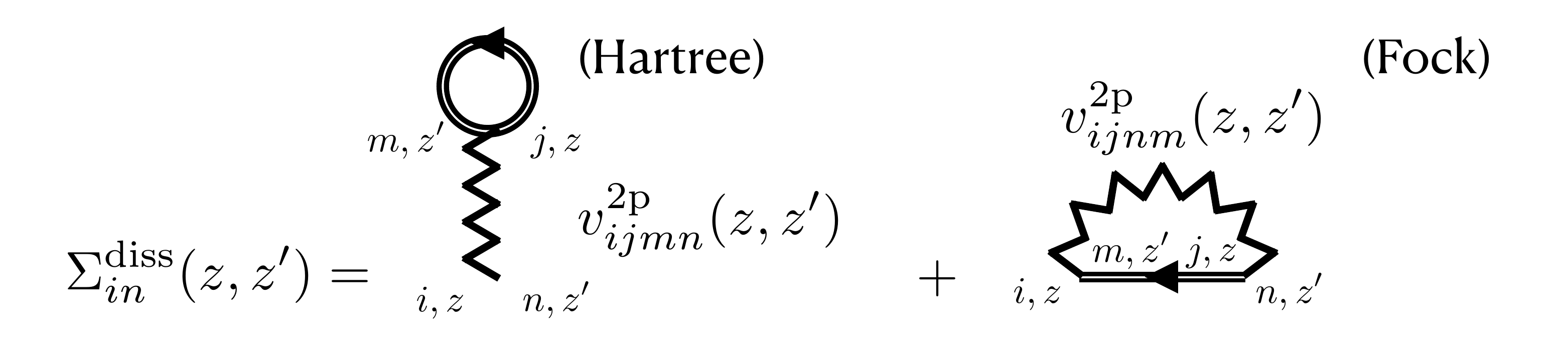}
\caption{Dissipation-induced self-energy diagrams -- oriented double lines represent $G$ 
and zigzag lines represent $v^{\rm 2p}$. 
The $(\pm)$ prefactor of the Hartree diagram can be 
reabsorbed as $\pm v^{\rm 2p}_{ijmn}=v^{\rm 2p}_{ijnm}$, hence 
$\S^{\rm H}=\S^{\rm F}$. 
It is readily seen that this is property holds true at any order.}
\label{Sigmaexamples}
\end{figure}

The second term in Eq.~(\ref{Hzatez2bd}) is quartic in the field 
operators and can be treated perturbatively, 
leading again to a Dyson equation. 
Unlike physical, e.g., Coulomb, interactions the contour-times $z$ and $z'$ 
are shared by two creation ($\hat{d}^{\dagger}\hat{d}^{\dagger}$)
and annihilation ($\hat{d}\hat{d}$) operators rather than by 
particle-hole-like operators ($\hat{d}^{\dagger}\hat{d}$).
This fact gives rise to slightly different 
self-energy diagrams,  with examples provided in 
Fig.~\ref{Sigmaexamples}. This difference is crucial to show that 
the number of topological equivalent 
diagrams of order $k$ is the same as for Coulomb-like interactions, 
i.e., $2^{k}k!$, despite
$v^{\rm 2p}(z,z')$ is not symmetric under the exchange 
$z\leftrightarrow z'$. Thus the 
prefactors of the Feynman diagrams are the same as in ordinary 
many-body perturbation theory.
Taking into account the (anti)symmetry properties of $v^{\rm 2p}$, 
the Hartree (tadpole) and Fock (oyster) 
diagrams in Fig.~\ref{Sigmaexamples}
yield the same result, namely $\S^{\rm H}=
\S^{\rm F}=\frac{1}{2}\S^{\rm HF}$ with
\begin{align}
\S^{\rm HF}_{in}(z,z')&=\mp 2iv^{\rm 
2p}_{ijmn}(z,z')G_{mj}(z',z)
\nn\\&=-2i[s(z)\d(z',z)+2\th_{+}(z)\d(z',\atez)]
v^{\rm 2p}_{ijmn}(t)\r^{<}_{mj}(t). 
\label{HFdiss}
\end{align}

At the Hartree-Fock (HF) mean-field level the r.h.s. of the
eom Eq.~(\ref{noninteom}) is modified by the addition 
of  $\int_{C}d\bar{z}\,\S^{\rm 
HF}_{in}(z,\bar{z})G_{np}(\bar{z},z')=-2v^{\rm 2p}_{ijmn}(t)\r^{<}_{mj}(t)
[is(z)G_{np}(z,z')+2i\th_{-}(\atez)G_{np}(\atez,z')]$. Remarkably, 
this term renormalizes $\ell^{>}$ in 
Eqs.~(\ref{eomGkc}) according to $\ell^{>}_{in}\to 
\ell^{>}_{in}+2v^{\rm 2p}_{ijmn}\r^{<}_{mj}$, while it leaves $\ell^{<}$ 
unchanged. Such asymmetry is due to the absence of two-body gain. 
Had we included Lindblad operators of the form  
$\hat{L}_{\g}=b^{\g\ast}_{mn}\hat{d}^{\dagger}_{n}\hat{d}^{\dagger}_{m}$ 
we would have found a similar renormalization for $\ell^{<}$. We 
infer that the mean-field treatment of two-body loss and gain 
is equivalent to considering one-body loss and gain. 

{\em Particle-hole loss.--}
The treatment of mixed Lindblad operators, 
containing both $\hat{d}$ and $\hat{d}^{\dagger}$, deserves a separate 
discussion, but it does not pose a conceptual 
problem~\cite{normalordering}.
 Let us consider 
the set 
$\hat{L}_{\g}=a^{\g}_{mn}\hat{d}^{\dagger}_{m}\hat{d}_{n}$ 
(particle-hole loss). 
As these dissipators are relevant in the context of phonon-induced 
relaxation of hot electrons in 
solids~\cite{taj_microscopic_2009,rosati_derivation_2014} we focus on 
the fermionic case. 
The normal-ordered form of the operator in Eq.~(\ref{hzatez}) reads
\begin{align}
\hat{H}(\bar{z},\bar{\atez})&=
\hat{H}(\bar{z})-is(z)V^{\rm ph}_{mn}(t)\hat{d}^{\dagger}_{m}(z)\hat{d}_{n}(z)
\nn\\&+\frac{1}{2}
\int_{C}d\bar{z}'v^{\rm ph}_{ijmn}(\bar{z},\bar{z}')
\hat{d}^{\dagger}_{i}(\bar{z}'^{+})\hat{d}^{\dagger}_{j}(\bar{z}^{+})
\hat{d}_{m}(\bar{z})\hat{d}_{n}(\bar{z}'),
\label{Hzatezph}
\end{align}
where $v^{\rm ph}_{ijmn}(z,z')=v^{\rm ph}_{ijmn}(t)
[-is(z)\d(z',z)+2i\th_{-}(z)\d(z',\atez)]$, with $v^{\rm 
ph}_{ijmn}(t)=2a^{\g\ast}_{ni}(t)a^{\g}_{jm}(t)$, and $V^{\rm ph}_{mn}(t)=\frac{1}{2}v^{\rm 
ph}_{mjnj}(t)$. The $V^{\rm ph}$-term renormalizes the one-particle 
Hamiltonian in Eqs.~(\ref{eomGkc}) according to $\tilde{h}(z)\to 
\tilde{h}(z)-is(z)V^{\rm ph}(t)$, but it does not renormalize the function 
$\ell(z)$. We now show that the mathematical structure of the dissipative 
KBE is recovered when treating $v^{\rm ph}$ at the HF mean-field level. 

For simplicity we take 
$a^{\g\ast}_{mn}=a^{\g}_{nm}$.
The diagrams with $v^{\rm ph}$ interaction-lines
are standard since the contour times $z$ and $z'$ are shared by particle-hole-like 
operators~\cite{precaution}. 
The Hartree (tadpole) diagram is easily shown to vanish 
whereas the Fock diagram yields
\begin{align}
\S^{\rm F}_{in}(z,z')&=\frac{i}{2}[v^{\rm ph}_{ijnm}(z,z')
+v^{\rm ph}_{ijnm}(z',z)]G_{mj}(z,z')
\nn\\
&=s(z)v^{\rm ph}_{ijnm}(t)[\d(z',z)-\d(z',\atez)]G_{mj}(z,z').
\end{align}
The r.h.s. of Eq.~(\ref{kb<tp}) then reads $[\S^{\rm F}\cdot 
G]_{ij}^{<}(t,t')=2i W^{<}_{in}(t)G^{\rm A}_{nj}(t,t')$ where 
$W^{<}_{in}(t)=\frac{1}{2}v^{\rm ph}_{ipnq}(t)\r^{<}_{qp}(t)$. This term, 
together with the $V^{\rm ph}$-renormalization of $\tilde{h}$ leads to 
a noninteracting dissipative KBE for $G^{<}$ 
with one-body loss and gain renormalized according to 
$\ell^{\lessgtr}\to \ell^{\lessgtr}+W^{\lessgtr}$, where 
$W^{>}=V^{\rm ph}-W^{<}$. Similarly, the r.h.s. of Eq.~(\ref{kb>tp}) yields
$[G\cdot\S^{\rm F}]^{>}_{ij}(t,t')=-2iG^{\rm 
R}_{in}(t,t')W^{>}_{nj}(t')+2iG^{>}_{in}(t,t')V^{\rm ph}_{nj}(t')$. Taking 
into account the $V^{\rm ph}$-renormalization of $\tilde{h}$ we find 
a noninteracting dissipative KBE for $G^{>}$ with the {\em same} 
renormalized $\ell^{\lessgtr}$ as for $G^{<}$. Once again,  
although through a different path, the mean-field treatment reduces 
the problem to considering one-body loss and gain.

{\em Beyond mean-field.--}
The self-energy diagrams do, in general, contain both physical 
and dissipation-induced interaction lines. 
To derive the KBE we first need to 
inspect the structure of the total self-energy
as a correlator on the Keldysh contour. In Appendix~\ref{selfapp} we show that
 the total self-energy can be written as $\S=\S^{\rm HF}+\S^{\rm 
corr}$, where $\S^{\rm HF}$ is the sum of all HF contributions 
and $\S^{\rm corr}$ satisfies the Keldysh properties.  
Therefore the Langreth rules 
remain unchanged and the KBE are still given by Eqs.~(\ref{kb<tp}) 
and (\ref{kb>tp}) with mean-field 
renormalized $\ell^{\lessgtr}$ and with $\S\to \S^{\rm corr}$.

{\em Conclusions.--} The second-quantization approach of 
NEGF theory has been extended to dissipative Lindbladian dynamics. We 
have shown how to derive the MSH for the $n$-particle NEGF and 
established the 
KBE for $G^{\lessgtr}$. We have generalized the diagrammatic rules for 
approximate treatments, derived the eom at the mean-field level, 
and elucidated the structure of the 
total self-energy as a correlator on the Keldysh contour. 
We hope that our contribution inspire further 
developments in the theory of many-body dissipative dynamics and 
stimulate first-principles studies of driven correlated open systems.

This work has been supported by 
MIUR PRIN (Grant No. 2022WZ8LME) and INFN through the TIME2QUEST 
project.

\appendix

\begin{widetext}

\section{Equations of motion}
\label{eomapp}

Let $\hat{O}_{i}$ be operators on the Keldysh contour $C$ and consider 
the contour-ordered string of operators 
\begin{align}
\hat{k} (z_1, \ldots,z_n) = \callT \left\{ \hat{O}_{1} (z_{1}) \ldots 
\hat{O}_{n} (z_{n}) \right\},  
\label{correlatorop}
\end{align}
where the contour argument of the operators 
fixes the 
position of these operators along the contour, thus rendering 
unambiguous the action of the contour ordering $\callT$.
The contour derivative of $\hat{k}$ with respect to the contour-time 
$z_{k}$ is given by~\cite{svl-book} 
\begin{align}
\frac{d}{dz_{k}} \hat{k} (z_1, \ldots,z_n)&= 
\sum_{l\neq k} (-)^{f_{kl}} \delta (z_k,z_l)  
 \callT\left\{\! \hat{O}_{1} \ldots 
 \left[ \hat{O}_k ,\hat{O}_l \right]_{\mp} \ldots 
 \hat{O}_{n} \! \right\},
 \label{genthetaderiv}
\end{align}
where $f_{kl}$ is the number of interchanges of fermionic operators 
required to bring $\hat{O}_l$ to the right of $\hat{O}_k$ (if 
$\hat{O}_l$ is a bosonic operator then $f_{kl}=0$ by definition), and 
the lower sign applies when {\em both} $\hat{O}_{k}$ and 
$\hat{O}_{l}$ are fermionic operators.
For a system with general Lindblad operators the NEGF is defined as, 
see Eq.~(\ref{klnegfdefmbp}), 
\begin{align}
G_{ij}(z,z')=\frac{1}{i}\Tr\Big[\hat{\r}(0)
\callT\Big\{e^{-i\int_{C}d\bar{z}\hat{H}(\bar{z},\bar{\atez})
}\hat{d}_{i}(z)\hat{d}^{\dag}_{j}(z')\Big\}\Big]\equiv 
\frac{1}{i}\left\langle \hat{d}_{i}(z)\hat{d}^{\dag}_{j}(z')
\right\rangle,
\label{klnegfdefmbpapp}
\end{align}	
where the operator $\hat{H}(\bar{z},\bar{\atez})$ is given in 
Eq.~(\ref{hzatez}).
Expanding the exponential of Eq.~(\ref{klnegfdefmbpapp}) in Taylor series and using 
Eq.~(\ref{genthetaderiv}) we find
\begin{align}
i\frac{d}{dz}G_{ij}(z,z')&=\d(z,z')+\frac{1}{i}
\int_{C}dz_{1}\d(z,z_{1})
\left\langle 
\left[\hat{d}_{i}(z),\hat{H}(z)-is(z)\hat{L}^{\dag}_{\g}(z)
\hat{L}_{\g}(z)\right]_{-}\hat{d}^{\dag}_{j}(z')\right\rangle
\nn\\
&+\frac{1}{i}\int_{C}dz_{1}\d(z,\atez_{1})2i\th_{-}(z_{1})
\left\langle\left[\hat{d}_{i}(z),\hat{L}^{\dag}_{\g}(z)\right]_{\mp}
\hat{L}_{\g}(z_{1})\hat{d}^{\dag}_{j}(z')\right\rangle
\nn\\
&+
\frac{1}{i}\int_{C}dz_{1}
(\pm) 2i\th_{-}(z_{1})\d(z,z_{1})\left\langle\hat{L}^{\dag}_{\g}(\atez_{1})
\left[\hat{d}_{i}(z),\hat{L}_{\g}(z)\right]_{\mp}\hat{d}^{\dag}_{j}(z')\right\rangle
\nn\\
&=\d(z,z')+\frac{1}{i}\left\langle 
\left[\hat{d}_{i}(z),\hat{H}(z)-is(z)\hat{L}^{\dag}_{\g}(z)\hat{L}_{\g}(z)\right]_{-}
\hat{d}^{\dag}_{j}(z')\right\rangle
\nn\\&
-2\th_{+}(z)\left\langle\left[\hat{d}_{i}(z),\hat{L}^{\dag}_{\g}(z)\right]_{\mp}
\hat{L}_{\g}(\atez)\hat{d}^{\dag}_{j}(z')\right\rangle
\pm 2\th_{-}(z)\left\langle\hat{L}^{\dag}_{\g}(\atez)
\left[\hat{d}_{i}(z),\hat{L}_{\g}(z)\right]_{\mp}\hat{d}^{\dag}_{j}(z')\right\rangle,
\label{klnegfderprelpapp}
\end{align}	
where the lower sign applies when {\em both} 
$\hat{d}_{i}$ and $\hat{L}_{\g}$ are fermionic 
operators. In the last identity we use that
\begin{align}
\int_{\g}dz F(z,\atez)&=\int_{0}^{\iif}dt 
F(t_{-},t_{+})+\int_{\iif}^{0}dt F(t_{+},t_{-})=-\int_{0}^{\iif}dt 
F(t_{+},t_{-})-
\int_{\iif}^{0}dt F(t_{-},t_{+})
\nn\\
&=-\int_{\g}dz F(\atez,z)
\label{atezid}
\end{align}
for any function $F$.
Similarly, we have 
\begin{align}
-i\frac{d}{dz'}G_{ij}(z,z')&=\d(z,z')-\frac{1}{i}\int_{C}dz_{1}
\d(z',z_{1})\left\langle \hat{d}_{i}(z)
\left[\hat{d}^{\dag}_{j}(z'),\hat{H}(z')-is(z')\hat{L}^{\dag}_{\g}(z')\hat{L}_{\g}(z')\right]_{-}
\right\rangle
\nn\\
&-\frac{1}{i}\int_{C}dz_{1}\d(z',\atez_{1})2i\th_{-}(z_{1})\left\langle \hat{d}_{i}(z)
\left[\hat{d}^{\dag}_{j}(z'),\hat{L}^{\dag}_{\g}(z')\right]_{\pm}\hat{L}_{\g}(z_{1})\right\rangle
\nn\\
&-\frac{1}{i}\int_{C}dz_{1}\d(z',z_{1})(\pm)2i\th_{-}(z_{1})
\left\langle \hat{d}_{i}(z)\hat{L}^{\dag}_{\g}(\atez_{1})
\left[\hat{d}^{\dag}_{j}(z'),\hat{L}_{\g}(z')\right]_{\pm}\right\rangle
\nn\\
&=\d(z,z')-\frac{1}{i}\left\langle \hat{d}_{i}(z)
\left[\hat{d}^{\dag}_{j}(z'),\hat{H}(z')-is(z')\hat{L}^{\dag}_{\g}(z')\hat{L}_{\g}(z')\right]_{-}
\right\rangle
\nn\\
&-2\th_{+}(\atez')\left\langle \hat{d}_{i}(z)
\left[\hat{d}^{\dag}_{j}(z'),\hat{L}^{\dag}_{\g}(z')\right]_{\pm}\hat{L}_{\g}(\atez')\right\rangle
\mp 2\th_{-}(z')
\left\langle \hat{d}_{i}(z)\hat{L}^{\dag}_{\g}(\atez')
\left[\hat{d}^{\dag}_{j}(z'),\hat{L}_{\g}(z')\right]_{\pm}\right\rangle.
\label{dGdzprime}
\end{align}	

For quadratic 
self-adjoint Hamiltonians 
$\hat{H}(t)=\hat{H}_{0}(t)=h_{mn}(t)\hat{d}^{\dag}_{m}\hat{d}_{n}$ and 
linear Lindblad operators 
$\hat{L}_{1\g}(t)=a_{n}^{\g}(t)\,\hat{d}_{n}$ (one-body loss) 
and $\hat{L}_{2\g}(t)=b^{\g \ast}_{n}(t)\,\hat{d}^{\dag}_{n}$ 
(one-body gain) the (anti)commutation rules can be used to simplify  
Eqs.~(\ref{klnegfderprelpapp}) and (\ref{dGdzprime}), the final 
outcome being Eqs.~(\ref{eomGkc}).
These equations can be written in a more compact form  by 
introducing the two-time matrices
\begin{align}
\tilde{h}(z_{1},z_{2})=\d(z_{1},z_{2})\tilde{h}(z_{2})
+2i\d(z_{1},\atez_{2})\ell(z_{2}).
\end{align}
Then
\begin{subequations}
\begin{align}
i\frac{d}{dz}G(z;z')-
\int_{C}d\bar{z}\,\tilde{h}(z,\bar{z})
G(\bar{z};z')=\d(z,z'),
\label{klnegfdermb2}
\\
-i\frac{d}{dz'}G(z;z')-
\int_{C}d\bar{z}\,G(z;\bar{z})\tilde{h}(\bar{z},z')
=\d(z,z').
\label{klnegfderadjmb2}
\end{align}	
\end{subequations}

\section{Wick's theorem}
\label{wickapp}

Let us define the $n$-particle NEGF according to 
\begin{align}
G_{n}(1,..,n;1',..,n')=\frac{1}{i}\left\langle
\hat{d}(1)..\hat{d}(n)\hat{d}^{\dag}(n')..\hat{d}^{\dag}(1')\right\rangle,
\label{klnegfdefmbnp}
\end{align}	
where $\hat{d}(q)\equiv \hat{d}_{j_{q}}(z_{q})$. This means that 
$G(1;1')=G_{j_{1}j'_{1}}(z_{1},z'_{1})$. Following the same steps 
leading to the equations of motion for the one-particle NEGF we find
\begin{align}
i\frac{d}{dz_{q}}G_{n}(1,..,n;1',..,n')&-\sum_{k}
\int_{C}d\bar{z}\,\tilde{h}_{j_{q}k}(z_{q},\bar{z})
G_{n}(1,..,k\bar{z},..,n;1',..,n')
\nn\\
&=\sum_{l=1}^{n}(\pm)^{l+q}\d_{j_{q}j'_{l}}\d(z_{q},z'_{l})
G_{n-1}(1,..,\stackrel{\sqcap}{q},..,n;1',..,\stackrel{\sqcap}{l'},..,n'),
\label{klnegfdermb2np}
\end{align}	
\begin{align}
-i\frac{d}{dz'_{q}}G_{n}(1,..,n;1',..,n')&-\sum_{k}
\int_{C}d\bar{z}\,
G_{n}(1,..,n;1',..,k\bar{z},..,n')\tilde{h}_{j'_{q}k}(\bar{z},z'_{q})
\nn\\
&=\sum_{l=1}^{n}(\pm)^{l+q}\d_{j'_{q}j_{l}}\d(z'_{q},z_{l})
G_{n-1}(1,..,\stackrel{\sqcap}{l},..,n;1',..,\stackrel{\sqcap}{q'},..,n'),
\label{klnegfderadjmb2np}
\end{align}	
where the argument $k\bar{z}$ is placed at position $q$ in 
Eq.~(\ref{klnegfdermb2np}) and in position $q'$ in 
Eq.~(\ref{klnegfderadjmb2np}). We also introduce the notation 
according to which the argument underneath the symbol 
``~$\stackrel{\sqcap}{}$~'' 
is missing.
It is easy to verify the validity of Wick's theorem, i.e.,
\begin{align}
G_{n}(1,..,n;1',..,n')=
\left|
\begin{array}{ccc}
G(1;1') & \ldots & G(1,n')
\\
\vdots & \vdots & \vdots \\
G(n,1') & \ldots & G(n,n')
\end{array}
\right|_{\pm},
\end{align}
where the $\pm$ signifies permanent/determinant.
Expanding
$G_{n}$ along row $q$:
\begin{align}
G_{n}(1,..,n;1',..,n')=\sum_{l=1}^{n}(\pm)^{l+q}G_{j_{q}j'_{l}}(z_{q},z'_{l})
G_{n-1}(1,..,\stackrel{\sqcap}{q},..,n;1',..,\stackrel{\sqcap}{l'},..,n').
\end{align}
This expression satisfies Eq.~(\ref{klnegfdermb2np}) provided that 
$G_{j_{q}j'_{k}}(z_{q},z'_{k})$ satisfies Eq.~(\ref{klnegfdermb2}). 
Similarly, expanding along column $q'$ 
\begin{align}
G_{n}(1,..,n;1',..,n')=\sum_{l=1}^{n}(\pm)^{l+q}G_{j_{l}j'_{q}}(z_{l};z'_{q})
G_{n-1}(1,..,\stackrel{\sqcap}{l},..,n;1',..,\stackrel{\sqcap}{q'},..,n'),
\end{align}
which is easily shown to satisfy Eq.~(\ref{klnegfderadjmb2np})  
provided that $G_{j_{l}j'_{q}}(z_{l},z'_{q})$ satisfies 
Eq.~(\ref{klnegfderadjmb2}).

\section{Self-energy as a correlator on the Keldysh contour}
\label{selfapp}

The self-energy diagrams for interacting dissipative systems 
contain both physical and dissipation-induced
interaction lines. 
For simplicity we here 
address only two-body loss dissipators; the simultaneous presence of 
all type of interactions can be worked 
out along the same line of reasoning.

We consider the set of Lindblad operators (sum of repeated indices is 
implicit)
$\hat{L}_{\g}=\sum_{mn}a^{\g}_{mn}\hat{d}_{m}\hat{d}_{n}$, 
with $a^{\g}_{mn}=\pm a^{\g}_{nm}$ for bosons/fermions.
To derive the equations of motionwe need to evaluate the following 
commutators, see Eqs.~(\ref{klnegfderprelpapp}) 
and (\ref{dGdzprime}), 
\begin{align}
\left[\hat{d}_{i}(z),\hat{H}(z)\right]_{-}=h_{in}(z)\hat{d}_{n}(z),
\end{align}
\begin{align}
\left[\hat{d}_{i}(z),\hat{L}^{\dag}_{\g}(z)\hat{L}_{\g}(z)\right]_{-}
=
v^{\rm 
2p}_{iqmn}(t)\hat{d}^{\dag}_{q}(z)\hat{d}_{m}(z)\hat{d}_{n}(z)\equiv\hat{\Th}_{i}(z),
\end{align}
\begin{align}
\left[\hat{d}_{i}(z),\hat{L}^{\dag}_{\g}(z)\right]_{-}\hat{L}_{\g}(\atez)
=
v^{\rm 
2p}_{iqmn}(t)\hat{d}^{\dag}_{q}(z)\hat{d}_{m}(\atez)\hat{d}_{n}(\atez)
\equiv\hat{\D}_{i}(z).
\end{align}
Taking into account that 
$\left[\hat{d}_{i}(z),\hat{L}_{\g}(z)\right]_{-}=0$ we conclude that 
(sum over repeated indices is implicit)
\begin{align}
i\frac{d}{dz}G_{ij}(z,z')=\d(z,z')+h_{in}(z)G_{nj}(z,z')
-s(z)\left\langle \hat{\Th}_{i}(z)
\hat{d}^{\dag}_{j}(z')\right\rangle
-2\th_{+}(z)\left\langle \hat{\D}_{i}(z)
\hat{d}^{\dag}_{j}(z')\right\rangle.
\label{dGdz}
\end{align}
Similarly we can extract the adjoint equation of motion. Taking into 
account that
\begin{align}
\left[\hat{d}^{\dag}_{j}(z'),\hat{H}(z')\right]_{-}=-\hat{d}^{\dag}_{m}(z')h_{in}(z'),
\end{align}
\begin{align}
\left[\hat{d}^{\dag}_{j}(z'),\hat{L}^{\dag}_{\g}(z')\hat{L}_{\g}(z')\right]_{-}
=\hat{\Th}^{\dag}_{j}(z'),
\end{align}
\begin{align}
\hat{L}^{\dag}_{\g}(\atez')\left[\hat{d}^{\dag}_{j}(z'),\hat{L}_{\g}(z')\right]_{-}
=\hat{\D}^{\dag}_{j}(z'),
\end{align}
and that 
$\left[\hat{d}^{\dag}_{j}(z),\hat{L}^{\dag}_{\g}(z)\right]_{-}=0$ we conclude that
\begin{align}
-i\frac{d}{dz'}G_{ij}(z,z')=\d(z,z')+G_{im}(z,z')h_{mj}(z')
+s(z')\left\langle \hat{d}_{i}(z)\hat{\Th}^{\dag}_{j}(z')
\right\rangle-2\th_{-}(z')\left\langle \hat{d}_{i}(z)\hat{\D}^{\dag}_{j}(z')
\right\rangle.
\label{eomzp2bdiss}
\end{align}	

Following the same steps we can calculate the derivative of 
$\left\langle \hat{\Th}_{i}(z)\hat{d}^{\dag}_{j}(z')\right\rangle$
and $\left\langle \hat{\D}_{i}(z)
\hat{d}^{\dag}_{j}(z')\right\rangle$
with respect to $z'$. We see that in Eq.~(\ref{dGdzprime})
the operator $\hat{d}_{i}(z)$ is a spectator. The result is therefore 
identical to Eq.~(\ref{eomzp2bdiss}) with $\hat{d}_{i}(z)\to 
\hat{\Th}_{i}(z)$ and $\hat{d}_{i}(z)\to 
\hat{\D}_{i}(z)$, respectively, and the delta function is replaced 
by $\d(z,z')\left\langle 
\left[\hat{\Th}_{i}(z),\hat{d}^{\dag}_{j}(z)\right]_{\mp}\right\rangle$
and $\d(\atez,z')\left\langle 
\left[\hat{\D}_{i}(z),\hat{d}^{\dag}_{j}(z)\right]_{\mp}\right\rangle$,  respectively.
Let us introduce the general correlator on the contour
\begin{align}
G^{AB}_{ij}(z,z')&=\frac{1}{i}\left\langle 
\hat{A}_{i}(z)\hat{B}^{\dag}_{j}(z')\right\rangle.
\end{align}
Then
\begin{align}
-i\frac{d}{dz'}G^{\Th d}_{ij}(z,z')=
\d(z,z')\left\langle 
\left[\hat{\Th}_{i}(z'),\hat{d}^{\dag}_{j}(z')\right]_{\mp}\right\rangle
+G^{\Th d}_{ij}(z,z') h_{mj}(z')
+i s(z')G^{\Th\Th}_{ij}(z,z')
-2i\th_{-}(z')G^{\Th\D}_{ij}(z,z'),
\label{dGTddzp}
\end{align}
\begin{align}
-i\frac{d}{dz'}G^{\D d}_{ij}(z,z')=
\d(\atez,z')\left\langle \left[\hat{\D}_{i}(\atez'),\hat{d}^{\dag}_{j}(z')\right]_{\mp}\right\rangle
+G^{\D d}_{ij}(z,z') h_{mj}(z')
+i s(z')G^{\D\Th}_{ij}(z,z')
-2i\th_{-}(z')G^{\D\D}_{ij}(z,z').
\label{dGDddzp}
\end{align}
In matrix form Eqs.~(\ref{dGdz}), (\ref{eomzp2bdiss}), 
(\ref{dGTddzp}) and (\ref{dGDddzp}) read
\begin{subequations}
\begin{align}
\big[i\frac{d}{dz}-h(z)\big]G(z,z')&=\d(z,z')
-is(z)G^{\Th d}(z,z')
-2i\th_{+}(z)G^{\D d}(z,z'),
\label{171}
\\
G(z,z')\big[-i\frac{\overleftarrow{d}}{dz'}-h(z')\big]&=\d(z,z')
+is(z')G^{d\Th}(z,z')
-2i\th_{-}(z')G^{d\D}(z,z'),
\\
G^{\Th d}(z,z')\big[-i\frac{\overleftarrow{d}}{dz'}-h(z')\big]&=\d(z,z') 
C^{\Th d}(z')+i s(z')G^{\Th\Th}(z,z')
-2i\th_{-}(z')G^{\Th\D}(z,z'),
\\
G^{\D d}(z,z')\big[-i\frac{\overleftarrow{d}}{dz'}-h(z')\big]&=
\d(\atez,z')C^{\D d}(z')
+i s(z')G^{\D\Th}(z,z')
-2i\th_{-}(z')G^{\D\D}(z,z'),
\end{align}
\end{subequations}
where 
\begin{align}
C^{\Th d}_{ij}(z)&\equiv \left\langle 
\left[\hat{\Th}_{i}(z),\hat{d}^{\dag}_{j}(z)\right]_{\mp}\right\rangle
=
v^{\rm 2p}_{iqmn}(t)
\left\langle \left[\hat{d}^{\dag}_{q}(z)\hat{d}_{m}(z)\hat{d}_{n}(z),
\hat{d}^{\dag}_{j}(z)\right]_{\mp}\right\rangle =2v^{\rm 
2p}_{iqmj}(t)\r^{<}_{mq}(t),
\nn\\
C^{\D d}_{ij}(z)&\equiv \left\langle 
\left[\hat{\D}_{i}(\atez),\hat{d}^{\dag}_{j}(z)\right]_{\mp}\right\rangle
=
v^{\rm 2p}_{iqmn}(t)
\left\langle \left[\hat{d}^{\dag}_{q}(\atez)\hat{d}_{m}(z)\hat{d}_{n}(z),
\hat{d}^{\dag}_{j}(z)\right]_{\mp}\right\rangle =2v^{\rm 
2p}_{iqmj}(t)\left\langle \hat{d}^{\dag}_{q}(\atez)\hat{d}_{m}(z)\right\rangle.
\end{align}
From Eq.~(\ref{171}) we infer that
\begin{align}
\int d\bar{z}\;\S(z,\bar{z})G(\bar{z},z')=-is(z)G^{\Th d}(z,z')
-2i\th_{+}(z)G^{\D d}(z,z').
\end{align}
Therefore, acting with $\big[-i\frac{\overleftarrow{d}}{dz'}-h(z')\big]$ from the 
right we get
\begin{align}
\S(z,z')+\int_{C}d\bar{z}d\bar{z}'\S(z,\bar{z})G(\bar{z},\bar{z}')
\S(\bar{z}',z')&=-is(z)\Big[\d(z,z')C^{\Th d}(z')+i s(z')G^{\Th\Th}(z,z')
-2i\th_{-}(z')G^{\Th\D}(z,z')\Big]
\nn\\
&-2i\th_{+}(z)\Big[
\d(\atez,z')C^{\D d}(z')
+i s(z')G^{\D\Th}(z,z')
-2i\th_{-}(z')G^{\D\D}(z,z')\Big].
\end{align}
The contribution of the commutators is 
\begin{align}
\S^{\rm HF}_{ij}(z,z')&=-is(z)\d(z,z')2v^{\rm 
2p}_{iqmj}(t)\r_{mq}(t)-2i\th_{+}(z)\d(\atez,z')2v^{\rm 
2p}_{iqmj}(t)\r_{mq}(t)
\nn\\&=
-2iv^{\rm 
2p}_{iqmj}(t)\r_{mq}(t)\big[s(z)\d(z,z')+2\th_{+}(z)\d(\atez,z')\big],
\end{align}
which is exactly the Hartree-Fock contribution arising for the 
diagrammatic rules, see Eq.~(\ref{HFdiss}). Taking into account that $\S G\S$ contributes to 
the reducible self-energy 
we conclude that the correlation self-energy is given by
\begin{align}
\S^{\rm corr}(z,z')=\Big[s(z)s(z')G^{\Th\Th}(z,z')-2s(z)\th_{-}(z')G^{\Th\D}(z,z')
+2\th_{+}(z)s(z')G^{\D\Th}(z,z')-4\th_{+}(z)\th_{-}(z')
G^{\D\D}(z,z')\Big]_{\rm irr}
\end{align}
where the subscript `` irr " signifies the one-particle 
irreducible part of the correlators.

Let us prove that this object satisfies the Keldysh properties.
We assume $t>t'$ and therefore $G^{\Th\D}(z,z')=G^{\D\D}(z,z')$ and 
$G^{\D\Th}(z,z')=G^{\Th\Th}(z,z')$ since $\hat{K}\equiv \callT\Big\{e^{-i\int_{t_{-}}^{t_{+}}d\bar{z}
\hat{H}(\bar{z},\bar{\atez})}\Big\}=1$. Then setting $z=t_{-}$ and $z'=t'_{+}$ we get
\begin{align}
\S^{\rm corr}(t_{-},t'_{+})=
\Big[-G^{\Th\Th}(z,z')\Big]_{\rm irr}.
\end{align}
Let us verify that the result does not change choosing $z=t_{+}$ and $z'=t'_{+}$ . We 
get
\begin{align}
\S^{\rm corr}(t_{+},t'_{+})=\Big[G^{\Th\Th}(z,z')
-2G^{\Th\Th}(z,z')\Big]_{\rm irr}=\S^{\rm corr}(t_{-},t'_{+}).
\end{align}
Next we choose $z=t_{-}$ and $z'=t'_{-}$. We get
\begin{align}
\S^{\rm 
corr}(t_{-},t'_{-})=\Big[G^{\Th\Th}(z,z')-2G^{\D\D}(z,z')\Big]_{\rm irr}.
\end{align}
On the other hand if we had chosen $z=t_{+}$ and $z'=t'_{-}$ we would 
have got
\begin{align}
\S^{\rm corr}(t_{+},t'_{-})=\Big[-G^{\Th\Th}(z,z')+2G^{\D\D}(z,z')
+2G^{\Th\Th}(z,z')-4
G^{\D\D}(z,z')\Big]_{\rm irr}=\S^{\rm corr}(t_{-},t'_{-}).
\end{align}
A similar analysis can be carried out for $t<t'$.
In this case $G^{\Th\D}(z,z')=G^{\Th\Th}(z,z')$ and 
$G^{\D\Th}(z,z')=G^{\D\D}(z,z')$, and therefore
\begin{align}
\S^{\rm 
corr}(t_{-},t'_{-})=\Big[G^{\Th\Th}(z,z')-2G^{\Th\Th}(z,z')\Big]_{\rm irr},
\end{align}
\begin{align}
\S^{\rm corr}(t_{-},t'_{+})=\Big[-G^{\Th\Th}(z,z')
\Big]_{\rm irr}=\S^{\rm corr}(t_{-},t'_{-}),
\end{align}
\begin{align}
\S^{\rm corr}(t_{+},t'_{-})=\Big[-G^{\Th\Th}(z,z')+2G^{\Th\Th}(z,z')
+2G^{\D\D}(z,z')-4
G^{\D\D}(z,z')\Big]_{\rm irr},
\end{align}
\begin{align}
\S^{\rm corr}(t_{+},t'_{+})=
\Big[G^{\Th\Th}(z,z')
-2G^{\D\D}(z,z')\Big]_{\rm irr}=\S^{\rm corr}(t_{+},t'_{-}).
\end{align}
In conclusion
\begin{align}
\S^{\rm corr}(z,z')=\th(z,z')\Big[G^{\Th\Th}(z,z')
-2G^{\D\D}(z,z')\Big]_{\rm irr}-\th(z',z)\Big[G^{\Th\Th}(z,z')
\Big]_{\rm irr}.
\end{align}

\end{widetext}


\begin{thebibliography}{88}
\bibitem [{not()}]{notelk}%
  \BibitemOpen
  \href@noop {} {}\bibinfo {note} {Consider $\hat{Q}(t_{\rm max})\equiv
  \callT\Big\{ e^{-i \big(\int_{t_{-}}^{t_{\rm max}}+\int_{t_{\rm
  max}}^{t_+}\big) d\bar{z}\hat{H}(\bar{z},\bar{\atez})}\Big\}$. We have
  $\hat{Q}(t)=1$ and $dQ(t_{\rm max})/dt_{\rm max}=0$. Therefore $\hat{Q}
  =\lim_{t_{\rm max}\to\iif}\hat{Q}(t_{\rm max})=\hat{Q}=1$.}\BibitemShut
  {Stop}%
\bibitem [{pre()}]{precaution}%
  \BibitemOpen
  \href@noop {} {}\bibinfo {note} {The only precaution we need to take is in
  counting the topologically equivalent diagrams, as $v^{\rm ph}(z,z')$ is not
  symmetric under the exchange $z\leftrightarrow z'$. We can use standard
  diagrammatic prefactors provided that we evaluate the diagrams with the
  symmetrized form $\frac{1}{2}[v^{\rm ph}(z,z')+v^{\rm
  ph}(z',z)]$.}\BibitemShut {Stop}%
\end{thebibliography}

%

\end{document}